\documentclass{aa}
\input psfig.sty

\def\grau{$^{\circ}$}
\def\sec{$^{\prime\prime}$}
\def\min{$^{\prime}$}
\def\km{$\rm\, km\, s^{-1}$}
\def\zz{\mbox{$[Z/Z_{\sun}]$}}
\def\fe{\mbox{$[Fe/H]$}}
\def\ebv{\mbox{$E(B-V)$}}
\def\sw{\mbox{$\Sigma W$}}
\def\caH{Ca\,II H}
\def\caK{Ca\,II K}
\def\iraf{{\bf IRAF}}
\def\mr{metal-rich}
\def\mp{metal-poor}
\def\ia{intermediate-age}
\def\gc{globular cluster}
\def\gcs{globular clusters}

\begin{document}
\thesaurus{11.06.1 --- 11.09.1 --- 11.19.4}
\title{Metallicity distribution of bulge \gcs\ in M\,31}
\author{P. Jablonka\inst{1} \and E. Bica\inst{2} \and C. Bonatto\inst{2} 
\and T.J. Bridges{*}\inst{3} 
\and M. Langlois\inst{1} \and D. Carter\inst{4}}
\institute{URA 173 CNRS, DAEC, Observatoire de Paris, 92195, Meudon
Principal Cedex, France \and Universidade Federal do Rio Grande do
Sul, IF, Dept. de Astronomia, CP\,15051, Porto Alegre 91501--970, RS,
Brazil \and Royal Greenwich Observatory, Madingley Road, Cambridge, UK
\and Astrophysics group, Liverpool John Moores University, Byrom
Street, Liverpool, L3 3AF, United Kingdom}
\thanks{Visiting Astronomer, William Herschel Telescope.  The WHT is
operated on the island of La Palma by the Royal Greenwich Observatory
at the Spanish Observatorio del Roque de los Muchachos of the 
Instituto de Astrofisica de Canarias.}
\offprints{P. Jablonka}
\date{Received; accepted}
\maketitle

\markboth{M\,31 bulge \gcs}{ }

\begin{abstract}

This paper presents 3600---5400\,\AA\ integrated spectra of 19 \gcs\ 
(or candidates) projected on the central regions of M\,31, $r\leq5.3$\min\
($\approx$1.2\,kpc). We check the cluster nature of these objects, and derive 
their ages, metallicities and reddenings.  From the initial sample, 16 objects 
turn out to be true star clusters, two are Galactic dwarf stars, and one 
is a high redshift background galaxy. Only two clusters are found to be
super \mr, suggesting that this phenomenon is not very
common. For some clusters, we cannot rule out the possibility that
they are of intermediate age; this requires confirmation by
observations at the calcium triplet. We also present the metallicity
histogram of this central bulge sample and discuss possible scenarios
to explain its properties.


\end{abstract}

\section{Introduction}

Globular clusters in the Galaxy and in the Local Group play a fundamental role 
in connecting studies of individual stars and integrated properties
of star clusters.

In this respect, studies of star clusters in the Andromeda galaxy (M\,31) are
particularly suitable. The proximity and inclination of this giant
spiral allow one  to study in detail each of its constituent
sub-systems, halo, disk, and bulge even near the galaxy center.

M\,31 has been already extensively surveyed for \gcs\ e.g., Vetesnik
(1962), Sargent et al. (1977) and the Bologna group catalogue by
Battistini et al. (1987).  The brightest clusters have been explored
by means of integrated spectroscopy, which makes it possible to
determine some of their fundamental parameters, such as metallicity,
age and reddening, as well as to confirm their true nature as \gcs\
(e.g. Burstein et al. 1984, Tripicco 1989, Huchra, Brodie \& Kent
1991, Jablonka, Alloin \& Bica 1992, Bica et al. 1992, and Huchra et
al. 1996).  A number of studies have concentrated on identifying
clusters near the center of M31, e.g., Alloin, Pelat \& Bijaoui
(1976), Auri\`ere, Coupinot \& Hecquet (1992) and Battistini et
al. (1993).

In the present paper, we concentrate on the central 10\min\ by 10\min\
region of M\,31 and derive velocities, metallicities, ages and
reddenings of the confirmed star clusters. This work is a first
attempt at getting a picture of the properties of \gcs\ in the inner
bulge of a spiral galaxy (other than our own), which in turn must help
constrain models of galaxy formation and evolution. One of our
principal goals is to determine how frequent the occurrence of super
\mr\ \gcs\ can be. Bulge central regions, locations of intense star
formation, appear to be one of the best places to find them.

The methods employed in the present work, i.e. comparison with
template cluster spectra and grids of spectral properties as a
function of age and metallicity, have been developed and extensively
used to study star clusters in our Galaxy (e.g. Bica \& Alloin
1986a,b), the Magellanic Clouds (e.g. Santos et al. 1995a), M\,31
itself (Jablonka et al. 1992), and more recently NGC\,5128 (Jablonka
et al. 1996).

This paper is organised as follows: Section\,2 gives a description of
the sample; Section\,3 details observations and reductions; Section\,4
deals with the radial velocity measurements; Section\,5 presents
measurements of equivalent widths and results for cluster parameters;
a discussion of the results is given in Section~6. Finally, we
summarise our conclusions in Section\,7.

\section{Description of the sample}

\begin{table*}[t]
\caption{General data}
\label{tab_GD}
\renewcommand{\tabcolsep}{5.20mm}
\begin{tabular}{lccllccl}
\hline\hline
NB&$\alpha(1950)$&$\delta(1950)$&\ $V_R$ (H91)&\ \ $V_R$ (TP)&V&Class&Other names\\
  &$h\ m\ s$&\grau\ \min\ \sec\ \ &\km&\ \ \km&mag\\
\hline                                               
1&0 39 47.1&41 03 13&-311$\pm$30&-345$\pm$35&15.79&A&V\,80, G\,169, Bo\,107\\
2&0 39 42.0&41 02 48&-310$\pm$50&-246$\pm$22&16.54&A&V\,78, G\,158, Bo\,96\\
3&0 39 45.6&41 01 31&-323$\pm$22&-331$\pm$11&15.20&A&V\,75, G\,165, Bo\,103\\  
4&0 39 49.1&41 01 16&-286$\pm$34&-200$\pm$25&16.33&A&V\,77, G\,174, Bo\,112\\
5&0 39 45.8&41 00 59&&+120$\pm$42&17.40&A&Bo\,104  \\
6&0 39 51.4&40 58 08&&+191$\pm$62&16.38&A&Bo\,118  \\
7&0 39 50.3&40 57 36&-497$\pm$29&-589$\pm$33&16.05&A&V\,67, G\,177, Bo\,115\\
8&0 39 50.2&40 56 18&&-196$\pm$56&17.13&A&G\,175, Bo\,114 \\
9&0 39 46.9&40 55 52&-90$\pm$43&-98$\pm$31&16.14&A&V\,61, G\,168, Bo\,106  \\
10&0 39 57.2&40 58 57&&-75$\pm$22&14.71&A&Bo\,124  \\
11&0 40 00.3&40 58 15&-483$\pm$25&-445$\pm$17&14.47&A&G\,185, Bo\,127\\
12&0 39 59.6&40 56 17&&-193$\pm$35&17.09&A&V\,66, G\,184, Bo\,126\\
13&0 40 07.5&40 57 37&&-477$\pm$22&16.52&A&V\,71, G\,190, Bo\,134 \\
14&0 39 52.0&41 01 09&&-311$\pm$30&17.27&B&Bo\,119  \\
17&0 39 50.3&41 01 05&&-561$\pm$51&18.26&A& \\
21&0 39 53.8&40 59 32&&-773$\pm$40&17.69&B&  \\
22$^*$&0 39 42.2&41 03 31&&-208$\pm$97&18.41&C&  \\
32$^*$&0 39 42.9&41 00 43&&-170$\pm$60&18.82&C&  \\
38$^{**}$&0 39 53.4&40 57 01&&43\,847$\pm$81&&C\\
\hline
\end{tabular}
\renewcommand{\tabcolsep}{4.0mm}
\begin{list}{Table Notes.}
\item H91 = Huchra et al. (1991); TP = This paper. The remaining information 
is from Battistini et al. (1993). (*) - found to be a Galactic star (Sect.~4); 
(**) - found to be a high redshift galaxy (Sect.~4).
\end{list}
\end{table*}

\begin{figure}  
\vskip -2.0cm
\hskip -0.0cm
\psfig{figure=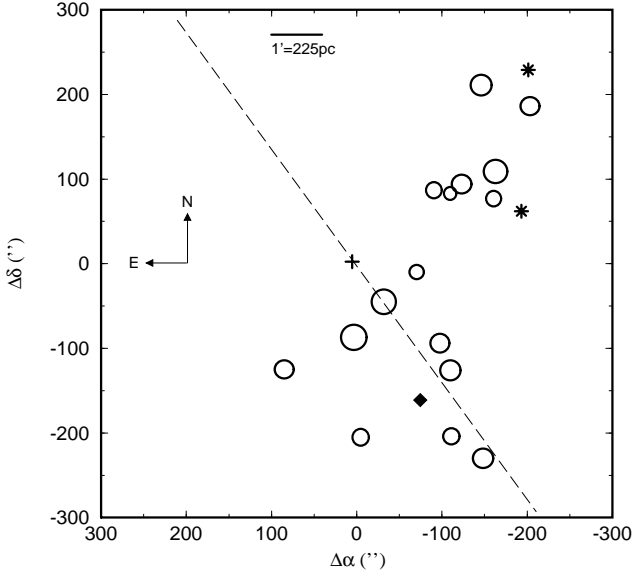,width=9.0cm,height=12.15cm}
\vskip -0.75cm
\caption[]{Angular distribution of the sample.  Star clusters are
shown by circles with radius proportional to the V brightness; the filled
diamond turned out to be a galaxy, and the asterisks Galactic dwarf
stars (Sect.~3). The dashed line indicates the orientation of the
major axis, and the plus sign the nucleus of M\,31.}
\label{fig1}
\end{figure}

\begin{figure}
\vskip 1.0cm
\hskip -0.0cm
\psfig{figure=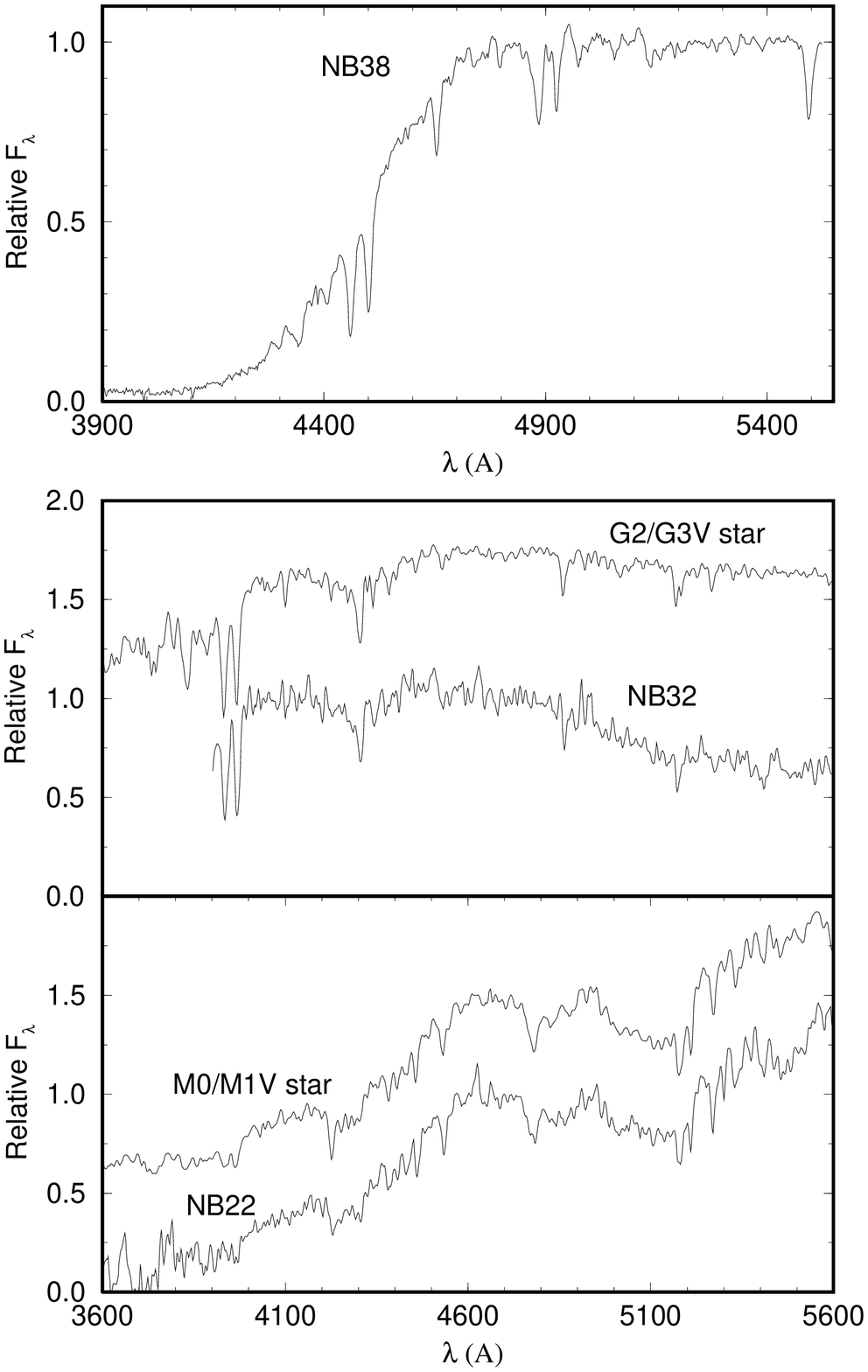,width=8.0cm,height=13.0cm}
\vskip -0.75cm
\caption[]{Top panel: the galaxy spectrum (NB\,38), not corrected for
redshift. Middle panel: the rest-frame spectrum of NB\,32 compared to
the stellar template of spectral types G2/G3\,V from Santos et al. (1995b); 
bottom panel: the same for the object NB\,22, compared to the star template 
M0/M1\,V. $F_{\lambda}$\ units are used and they are normalised at 
$\lambda\approx4570$\,\AA; the top spectra have been offset for clarity.} 
\label{fig2}
\end{figure}

We selected objects in the inner region of M\,31, either previously
confirmed star clusters or candidate clusters, from the recent list
published by Battistini et al.  (1993). Objects in their list with
probability class to be \gcs\ of A, B and C were considered in the
present study, with priority given to the brightest and reddest
ones. We optimised the positioning and orientation of the slits in the
masks in order to maximise the number of clusters observed in a single
exposure and ended up with 24 different objects.  Usable spectra have
been acquired for 19 of them, among which 12 have never been observed
spectroscopically. We list some useful information about our sample in
Table~1. 

The distribution of the objects on the sky is shown in Figure 1. The
objects span a region roughly 5' x 9' (1.1 x 2.0 kpc assuming a true
distance modulus of (m-Mo)=24.43; Freedman \& Madore 1990) just to the
west and south of the nucleus.  The position of the M\,31 nucleus and
the orientation of the major axis are also shown in the figure.

Table~\ref{tab_GD} contains, by columns: (1) - object number according
to Battistini et al.  (1993); (2) and (3) - right ascension and
declination (B1950); (4) - radial velocity given by Huchra et
al. (1991); (5) - radial velocity measured from our spectra (Sect.~3);
(6) V magnitude (Battistini et al. 1993); (7) - Bologna group
probability class for the object to be a \gc: A - very high, C - low;
and (8) - designations in other catalogues: V = Vetesnik (1962), G =
Sargent et al. (1977), and Bo = Battistini et al. (1987). Some
clusters have additional designations which are given in Battistini et
al. (1993).

\section{The observations}

The observations were carried out with the 4.2m William Herschel
Telescope on La Palma with the LDSS-2 multi-slit spectrograph for
which appropriate masks were designed with the LEXT program. A slit
width of 1.5\sec\ was used, and slit lengths varied between 10\sec ---
60\sec. Four fiducial stars were used for field acquisition. Two masks
have been created with 24 objects (Mask\,1) and 18 objects (Mask\,2)
respectively; 9 objects were in common between the two masks.

Spectroscopy of our two masks was obtained on October 19, 1995, using
a $1024\times1024$\ TEK CCD detector with $24\mu$m pixels. The image
scale is 0.59\sec/pixel and the field is $\sim$10\min$\times$
10\min. For each mask, we acquired exposures with the 2.4\AA\ /pixel
``high'' dispersion grism, centered on 4200\AA. The total wavelength
range is 3300---6100\AA, see Table~\ref{tab_OS} for further
details. The standard stars BD\,+28\grau4211, Feige\,110 and G191B2B
(Stone 1977, Oke 1974) were observed through a wide-slit for flux
calibration. Finally, CuAr arcs were taken throughout the night for
wavelength calibration.

\subsection{Data reduction}

We adapted the {\tt Multired} tool developed by Olivier Le F\`evre and
ran it under \iraf\ (Le F\`evre et al. 1995). {\tt Multired} allows
the simultaneous reduction of spectra in a given mask. It makes use of
the \iraf\ {\tt ccdred} and {\tt longslit} packages, in the usual
way. Although flat field frames were taken, they could not be used due
to some technical problems.  Consequently, we limited our analyses to
the 3600---5400\,\AA\ range, where the Signal/Noise ratio is higher,
and where the spectral distribution is not disturbed by border
effects. The final spectral resolution, as measured from the
comparison arcs, is $\sim$7\AA.

The cluster G\,185 (NB\,11) has a small companion 4\sec\ to the
North-West (Couture et al. 1995), which was not included in our
aperture.

The object NB\,38, which in Table~\ref{tab_GD} was Bologna class C,
turns out to be a high redshift galaxy, with $z \sim 0.13$. The
observed spectrum of this galaxy, not corrected for redshift, is shown
on the top panel of Figure~\ref{fig2}. A simple calculation shows that
a typical galaxy placed at the measured redshift would indeed exhibit
a size comparable to M31 genuine globular clusters. Its observation
points however towards a rather low extinction in M31 along this line
of sight.  NB\,22, also a class C object, is a dwarf Galactic star as
shown in the bottom panel of Figure~\ref{fig2}.  We compare this
spectrum with the stellar template of spectral type M0/M1\,V by Santos
et al.  (1995b). The agreement is excellent, which indicates that
NB\,22 is a foreground dwarf Galactic star. We also tested giant
stars, but spectral features, continuum distribution and luminosity
arguments rule out this possibility. The object NB\,32 (Bologna class
C) is also a Galactic dwarf star of spectral type G\,2/G\,3 (middle
panel), as indicated by the comparison with the corresponding template
from Santos et al. (1995b).

\begin{table}[h]
\caption{Observational Setup}
\label{tab_OS}
\renewcommand{\tabcolsep}{0.9mm}
\begin{tabular}{lcccc}
\hline\hline
 &\multicolumn{2}{c}{Mask\,1}&&\multicolumn{1}{c}{Mask\,2}\\
\hline                                               
Number of objects&&  24  &&  18             \\
Dispersion (\AA) && 2.4  &&  2.4            \\
Resolution (\AA) &&   6  &&  6              \\
Wavelength Coverage (\AA) && 3200---6100 && 3200---6100   \\
Total exposure time (sec) && 7200 &&  7200           \\
Individual exposures (sec) && 4$\times$1800 && 600+1200+3$\times$1800\\
\hline
\end{tabular}
\renewcommand{\tabcolsep}{4.0mm}
\end{table}

\section{Radial velocities}

The method we employed for the radial velocity determinations is the
cross-correlation of object against template spectra (Tonry \& Davis
1979). We formed a cross-correlation template using the 7 clusters
with known velocities from Huchra et al. (1991). These 7 cluster
spectra were continuum-normalised, shifted to zero velocity and
co-added using \iraf\ routines. In this resulting spectrum, we checked
strong absorption features (e.g. \caH\ and K, H Balmer lines and CH
G-band) to ensure that this template is indeed at zero velocity. As
well, we cross-correlated each of the 7 input spectra against the
template, and found a mean difference of 7\km\ with a standard
deviation of 60\km\ between the input and measured velocities.

Using the \iraf\ routine {\tt xcsao} within the {\tt rvsao} package,
the spectra of the objects have been cross-correlated with the
zero-velocity template, and the results are shown in column 5 of
Table~1. The uncertainties given in the Table are one sigma errors.
There is good agreement in the velocity values for the 7 \gcs\ in
common with Huchra et al. (1991). The blueshifts indicate membership
to M\,31. The 2 objects somewhat redshifted, NB\,5 and NB\,6, can be
considered as belonging to the tail of the velocity distribution of
M\,31 clusters (e.g. Huchra et al. 1991).

\begin{figure} 
\vskip 2.0cm
\hskip -0.0cm
\psfig{figure=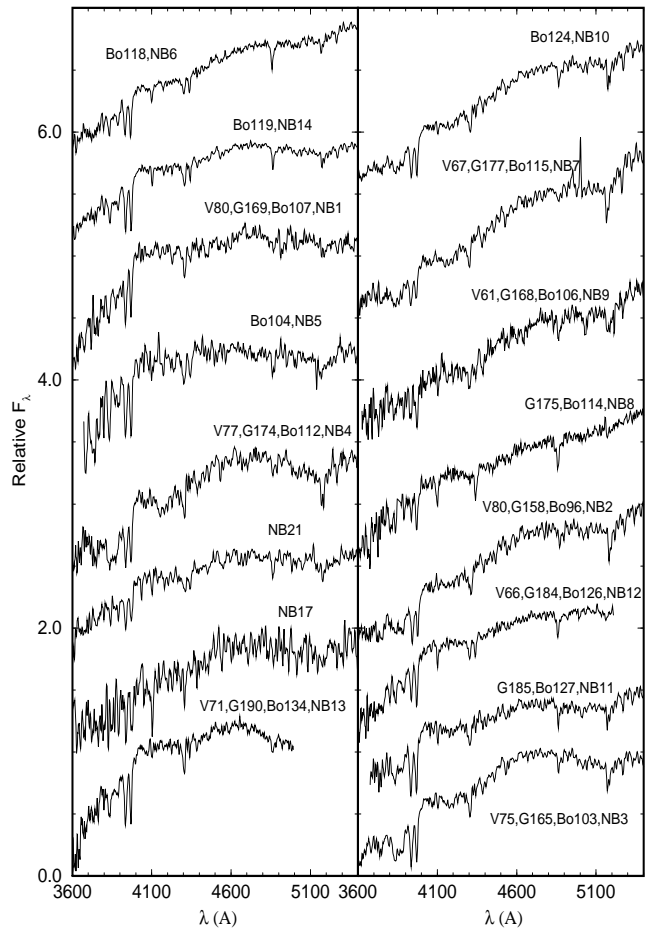,width=8.0cm,height=12.0cm}
\vskip -0.75cm
\caption[]{Spectra of the objects confirmed to be star clusters, 
rebinned to the rest-frame. Units as in Figure~\ref{fig2}. }
\label{fig3}
\end{figure}

\begin{figure} 
\vskip 2.0cm
\hskip -0.0cm
\psfig{figure=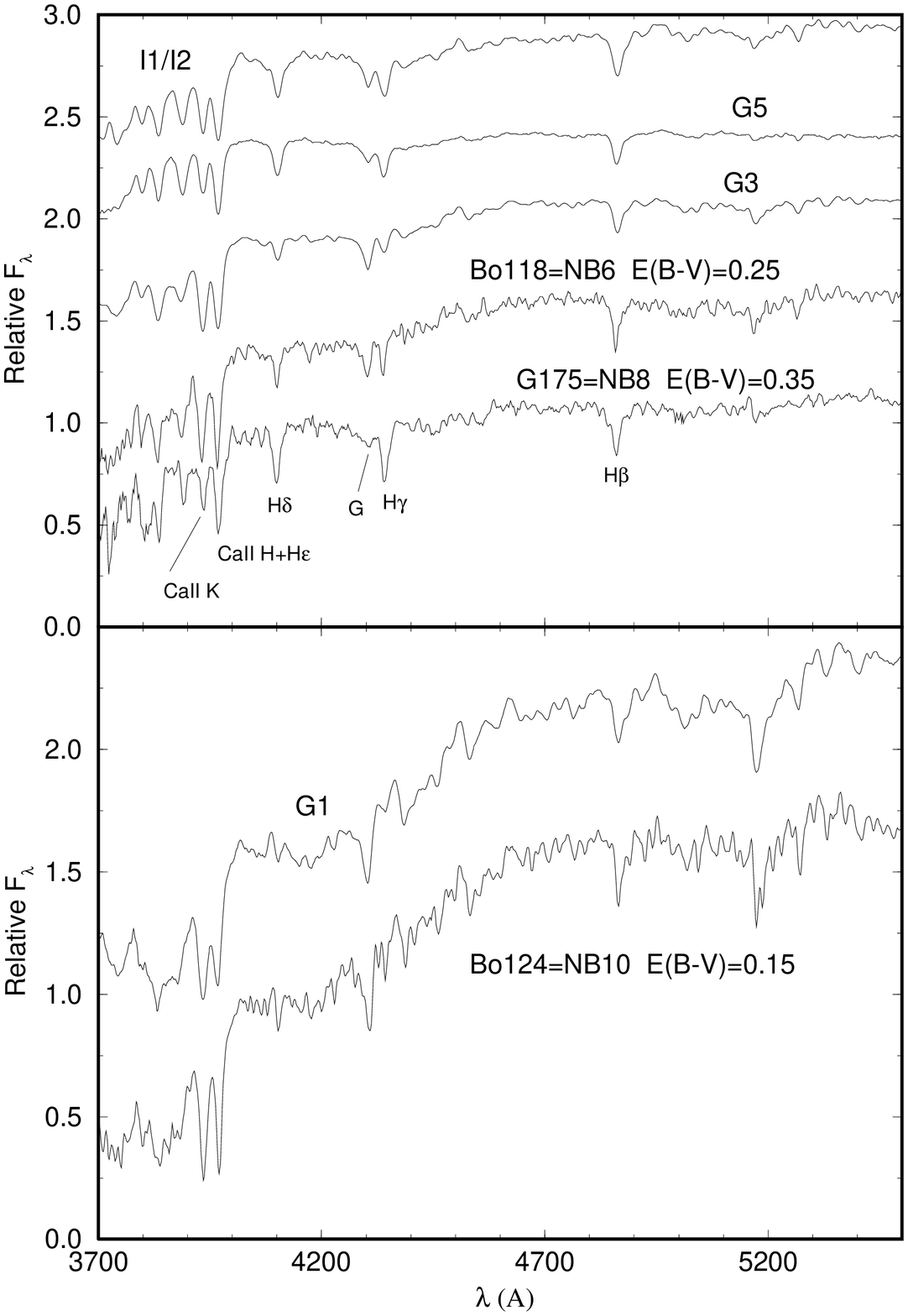,width=8.0cm,height=12.0cm}
\vskip -0.75cm
\caption[]{Method of reddening determination. Top panel: the spectrum
of cluster G\,175, corrected for $\ebv=0.35$, to match the continuum
distribution of either a \mp\ \gc, or the
\ia\ cluster template; same for Bo\,118 corrected for
$\ebv=0.25$, adopting the G\,3 template. Bottom panel: same as above for
the \mr\ \gc\ Bo\,124, corrected for $\ebv=0.15$, and
compared to the G\,1 template. Units as in Figure~\ref{fig2}.}
\label{fig4}
\end{figure}

\begin{table*}
\caption{Equivalent Widths in \AA}
\label{tab_EW}
\begin{tiny}
\renewcommand{\arraystretch}{1.25}
\renewcommand{\tabcolsep}{1.62mm}
\begin{tabular}{clc|rrrrrrrrrrrrrrr}
\hline\hline
Window& Absorber& Limits & G\,169&G\,158&G\,165&G\,174&Bo\,104&Bo\,118&G\,177&G\,175&G\,168&Bo\,124&G\,185&G\,184&G\,190&Bo\,119&\\
  & &(\AA)&NB\,1&NB\,2&NB\,3&NB\,4&NB\,5&NB\,6&NB\,7&NB\,8&NB\,9&NB\,10&NB\,11&NB\,12&NB\,13&NB\,14&NB\,21\\
\hline                                                               
  2&  CN, H9 &3814-3862  & 9.3 &19.3 &19.8 &25.5 & 7.9 & 8.5 &22.9&9.3&13.2:&17.5&11.6& 8.6&11.4&10.9&5.3\\
  3&  CN, H8 &3862-3908  &7.1  &15.0 &13.8 &18.8 & 8.6 & 7.7 &14.4&4.6&13.3: &11.8& 8.1& 8.6& 7.9& 7.7&6.1\\
  4&  \caK &3908-3952 &11.9&20.9 &18.8 &23.1 &13.2 & 10.6 &19.8&8.0&12.9&17.3&13.7& 9.7&15.3&14.1&10.2\\
  5&  \caH,H$\epsilon$&3952-3988&10.6&15.3&14.1&16.3&12.1&10.2&13.5&10.3&15.4&13.6&11.3&10.1&11.6&9.4&8.5\\
  9&  H$\delta$&4082-4124  &2.8 & 3.2 & 3.3 & 5.3 & 4.1: & 4.3 &4.4&5.5&5.0:& 4.3& 2.2& 4.6& 2.7& 2.2& 3.5\\
 10&  Fe\,I&4124-4150  &1.2 & 3.5 & 3.4 & 6.0 & 0.8: & 1.2 &5.2&0.3:&0.8:& 3.5& 2.9& 1.2:& 1.7& 2.0& 1.3\\
 11&  CN&4150-4214  &4.2 &10.3 &11.6 &17.8 & 3.2: & 4.7 &15.3&1.4:&8.9&11.5& 7.6& 2.6& 4.2& 5.6& 5.6\\
 12&  Ca\,I&4214-4244  &1.9 & 4.5 & 4.5 & 6.1 & 3.1: & 2.6 &5.2&1.6&3.4&4.3& 2.9& 2.2& 2.2& 3.9& 2.5\\
 13&  Fe\,I&4244-4284  &2.7 & 6.3 & 6.4 & 7.2 & 3.8: & 3.4 &7.2&2.7&7.2& 5.3& 3.8& 2.9& 4.2& 5.7& 3.8\\
 14&  CH G&4284-4318  &5.9 & 9.9 & 9.8 &10.7 & 6.2 & 5.7 &9.8&4.3&9.8& 8.6& 6.5& 5.7& 7.2& 8.5& 6.3\\
 15&  H$\gamma$&4318-4364  &4.7 & 4.2 & 6.2 & 4.9 & 4.4 & 5.6 &4.7&8.0&7.8& 5.3& 4.8& 5.0& 5.1& 6.6& 6.4\\
 16&  Fe\,I&4364-4420  &3.6 & 6.8 & 7.8 & 8.0 & 4.0: & 4.4 &8.6&3.1&7.7& 6.4& 5.5& 3.7& 6.2& 5.7& 3.8\\
 27&  H$\beta$&4846-4884  &3.8 & 3.5 & 3.7 & 3.3 & 2.7: & 4.0 &4.1&4.8&4.5& 3.7& 3.4& 3.8& 4.3& 5.4& 4.0\\
 31&  Fe\,I&4998-5064  &2.3 & 5.2 & 6.6 & 8.4 & 2.1: & 2.7 &5.0&4.0&6.8& 4.8& 3.8& 1.4:& --- & 4.2& 4.8:\\
 32&  Fe\,I, C$_2$&5064-5130  &2.6 & 5.1 & 6.5 & 9.8 & 4.3: & 3.1 &9.9&3.1&5.6& 5.0& 4.1& 1.7:&  ---& 1.8& 1.2:\\
 33&  MgH, C$_2$&5130-5156  &1.6 & 2.9 & 3.3 & 5.1 & 2.5: & 1.5 &4.6&1.1&2.5& 2.7& 2.0& 1.3:& --- &1.8& 2.5:\\
 34&  Mg\,I, MgH&5156-5196  &4.2 & 8.0 & 8.0 &11.9 & 5.3: & 4.0 &11.0&2.7&7.3& 7.0& 5.3& 2.7:& --- & 4.8& 6.0:\\
 35&  MgH&5196-5244  &2.6 & 4.6 & 5.0 & 6.4 & 4.0: & 2.4 &6.6&2.0&4.0& 4.4& 2.7& --- &---& 3.4& 4.0:\\
 36&  Fe\,I&5244-5314&2.4 & 4.0 & 4.5 & 5.6 & 2.4: & 2.6 &6.3&1.5&3.2& 4.6& 3.3& --- &---& 3.4& 2.6:\\
\hline
\end{tabular}
\renewcommand{\tabcolsep}{4.0mm}
\renewcommand{\arraystretch}{1.0}
\end{tiny}
\end{table*}

\section{Cluster parameters}

Figure~\ref{fig3} shows the resulting calibrated 
spectra rebinned to the rest frame 
and roughly arranged according to 
continuum slope. Notice that some clusters are quite
red but do not present strong 
metallic absorption lines, which suggests significant
reddenings. Also, metallic 
features show a wide range of strengths indicating a range 
in metallicities.

The equivalent widths (EWs) of the spectral absorption features were measured 
according to the window limits and continuum levels defined by Bica \& Alloin 
(1986a,b, 1988). The results are 
shown in Table~\ref{tab_EW} where we give the 
window number, main absorbers, window 
limits and provide the resulting EWs for the
sample objects. The objects NB\,22 and NB\,32 (Galactic dwarf stars) and NB\,38 
(a galaxy) are not included in the analysis. Also, the faintest star cluster in 
the sample, NB\,17 (Table~1), has been excluded 
because the Signal/Noise ratio of its
spectrum is too low for the present 
purpose (Fig.~\ref{fig3}). Typical equivalent 
width errors arise mainly from uncertainties 
in the continuum levels, and they imply 
a 5\% error for strong features 
as \caK, and a 10---15\% error for the weaker ones.

Two objects, G\,177 and G\,158, are in common with Jablonka et al. (1992). A 
comparison of EWs between the two works shows a very good agreement for G\,177, 
within 5\% for the strong absorption features, and 10---20\% for the weak ones. 
On the other hand, the values differ significantly for G\,158 in the sense that 
they are smaller in the present spectrum.  We 
suspect contamination or instrumental 
problems in the Jablonka et al. (1992) 
spectrum of G158.  Despite this, we confirm a
high metallicity (see below) and establish its nature as an old 
\gc\ rather than 
a possible \ia\ cluster (Bica et al. 1992).

The emission lines [OIII]$_{\lambda4959,5007}$\ 
and [OII]$_{\lambda3727}$, superimposed
on the spectrum of G\,177 (Figure~3), deserve 
special mention. In the spectrum of
Jablonka et al. (1992), in spite of the 
lower resolution, the [OIII]$_{\lambda5007}$\
line can be recognised, together 
with H$\alpha$\ which fills in the stellar absorption,
appearing in emission. In the present spectrum we measured a ratio
$\log$([OII]$_{\lambda3727}$/[OIII]$_{\lambda5007}$)=$-0.61$. This observed
ratio implies photoionisation, and 
any reddening correction would further confirm this
mechanism instead of 
shocks (Baldwin, Phillips \& Terlevich 1981). An inspection of the 
region of G\,177 in the Hodge (1981) M\,31 Atlas 
does not show any prominent H\,II region
associated with the M\,31 inner disk.  A possible 
explanation for this emission might be
diffuse gas ionised by hot post-AGB stars of this very \mr\ \gc\ itself,
similar to the scenario proposed 
by Binette et al. (1994) to explain gas emission associated with 
very \mr\ old populations in the 
central regions of galaxies. In fact, in the UV,
some moderately \mr\ Galactic \gcs\ have a flux excess shortward of
2000\,\AA\ (Bonatto, Bica \& Alloin 1995 and references therein).
An alternative to this hypothesis is the presence of planetary
nebula in the cluster.

\subsection{Ages and metallicities from Bica \& Alloin's (1986b) grid}

In order to determine ages and metallicities 
we first use a grid of star cluster 
spectral properties from Bica \& Alloin (1986b). This 
grid connects EWs with age and 
\zz. Following Jablonka et al. (1992), we use a 
minimisation procedure which compares  
the EWs of the present observed 
clusters (Table~3) with those of the grid. Although age 
and metallicity are given in relatively 
large steps in the grid, this method allows one 
to test the possibility that the star cluster is younger than classical \gcs.
The results for metallicity \zz\ and ages 
are given, respectively in columns 2 and 3 of
Table~\ref{tab_CP}.

For some clusters, the age/metallicity solution is not unique within the errors 
in EWs. Indeed, old \mp\ \gcs\ and \ia\ star clusters 
are nearly indistinguishable in the blue-violet 
region, e.g. Bica \& Alloin (1986a,b)
and Bica, Alloin \& Schmitt (1994). The 
blue Horizontal Branch in \mp\ \gcs\ mimics 
the relatively high main sequence and turnoff region of
\ia\ clusters (1---5\,Gyrs), producing similar spectral distributions and
enhanced Balmer absorption. In such 
cases, it is necessary to observe the clusters 
in the near-IR, particularly 
at the Ca\,II triplet in order to check for possible
dilution effects in the blue-violet (Bica \& Alloin 1987). The
region of the M\,31 bulge might also contain \ia\ clusters 
from the inner disk.

\begin{table*}
\caption{Cluster parameters}
\label{tab_CP}
\renewcommand{\tabcolsep}{2.5mm}
\begin{tabular}{lcccccrrcc}
\hline\hline
Name& \zz & age (Gyr)& Template& \ebv&\sw\ (\AA) &\zz&\fe&\fe\\
 & grid& grid  &   &  & metals &(*)& H91&CM94\\
\hline                                               
NB\,1=G\,169&$-$1.0&$\geq10$&G\,3&0.05$\pm$0.04& 31.4&$-$1.10& $-$1.18&$-$1.35\\
NB\,2=G\,158&0.0&$\geq10$&G\,1&0.10$\pm$0.05& 62.2&$+$0.03& $-0.26$\\
NB\,3=G\,165&0.0&$\geq10$& G\,1&0.04$\pm$0.04& 62.3&$+$0.04&$-$0.56&$-0.69$\\    
NB\,4=G\,174&0.6&$\geq10$&G\,1$^\dagger$&0.04$\pm$0.04&78.9&$+$0.55 &$+$0.29&$-$0.13\\
NB\,5=Bo\,104&$-$1.0& $\geq10$  &G\,3&0.00$\pm$0.03&34.3&$-$0.98 & \\
NB\,6=Bo\,118&$-$1.0& $\geq10$ & G\,3&0.25$\pm$0.03& 32.6&$-$1.05& \\
NB\,7=G\,177&0.4&$\geq10$&G\,1$^\dagger$&0.04$\pm$0.02&71.1&$+$0.31&$-$0.15&$-$0.32\\
NB\,8=G\,175&$-$1.0/$-$1.5&1---5 or $\geq10$?&I\,1/I\,2 or G\,3/G\,4?&0.35$\pm$0.05&21.4&$-$1.53 & &$-$1.16\\
NB\,9=G\,168&0.0/$-$0.5&$\geq10$&G\,1/G\,2& 0.17$\pm$0.07&46.6&$-$0.48&&$-$1.13\\
NB\,10=Bo\,124  &0.0&$\geq10$& G\,1&0.15$\pm$0.02& 56.9&$-$0.14& \\
NB\,11=G\,185  &$-$0.5&$\geq10$& G\,2&0.02$\pm$0.02 & 42.9&$-$0.63& $-$1.08&$-$1.19\\
NB\,12=G\,184&$-$1.5/$-$1.0&$\geq10$&G\,3/G\,4&0.12$\pm$0.02&22.3&$-$1.50&&$-$0.91\\
NB\,13=G\,190  &$-$0.5&$\geq10$&G\,2&0.02$\pm$0.03&41.0&$-$0.71&&$-$1.96\\
NB\,14=Bo\,119&$-$0.5&$\geq10$&G\,2&0.02$\pm$0.03&45.5&$-$0.53&\\
NB\,21&$-$1.0&$\geq10$&G\,3&0.03$\pm$0.03&33.5&$-$1.01\\
\hline
\end{tabular}
\renewcommand{\tabcolsep}{4.0mm}
\begin{list}{Table Notes.}
\item H91 = Huchra et al. (1991); CM94 = Cohen \& Matthews (1994). (*) this
metallicity is based on a relation for \gcs\ (Sect.~5.3). $\dagger$: this cluster 
is more \mr\ than the template G\,1, which was used for the reddening determination.
\end{list}
\end{table*}

The top panel of Figure~4 illustrates clearly the problem of
\ia/old \mp\ degeneracy. We display an average of the
\ia\ templates I\,1 and I\,2 (1 to 3\,Gyrs) from Bica
(1988), which includes Galactic open clusters and LMC clusters, and
the templates G\,5 and G\,3, which are averages of Galactic \gcs\ with 
$\zz\approx-2.0$\ and $-1.0$, respectively. Notice the
resemblance in Balmer lines and metallic features, as well as in
continuum distribution. We also show the M\,31 clusters Bo\,118 and
G\,175 to illustrate the difficulty in classifying unambiguously these
objects, especially G\,175, because of its unusually strong Balmer
lines. In Table~\ref{tab_CP} we point this out explicitly for
G\,175, and it must be kept in mind that a wider spectral range should
be observed in order to establish the nature of all objects with
$\zz\leq$-$1.0$. The occurrence of \ia\ clusters with metallicity
$\zz\approx-1.0$\ cannot be ruled out in a scenario of cannibalism
(Section~6.3). In fact, \ia\ clusters with this metallicity, such as NGC\,419 
in the SMC (Bica \& Alloin 1986a), could have been accreted.

Assuming that G\,175 and Bo\,118 are indeed \gcs, they
would show intrinsic differences compared with Galactic \gcs. At
a given metallicity, these two clusters have exceedingly strong Balmer
absorption, as previously reported by Burstein et al. (1984). This
can be verified with the help of EWs of metallic features
(Table~\ref{tab_EW}) and relative strengths of Balmer to nearby
metallic features, such as \caK\ {\it vs} H+H$\epsilon$, and G band
{\it vs} H$\gamma$. Bo\,118 has EWs of metallic features like those of
the template G\,3 ($\zz\approx$-$1.0$), but metallic/Balmer line ratios
like those of G\,4 ($\zz\approx$-$1.5$). G\,175 is further shifted with
respect to average properties of Galactic \gcs, in the
sense that metallic features are like those of G\,3/G\,4, and the
metallic/Balmer ratios are more similar to those of the G\,5 template
($\zz\approx-2.0$). This would suggest a more populated 
blue-Horizontal Branch in M\,31 clusters at metallicities comparable
to Galactic globular clusters.
We point out that in our Galaxy there
exists a considerable dispersion in the Horizontal Branch properties
of \gcs\ at a given metallicity, the second parameter
phenomenon, at least in the halo (e.g. Zinn 1980).

\subsection{Reddening from templates}

Ideally, the best spectroscopic method to determine reddening for star
clusters is to have available reddening-free template spectra of
similar properties, and to use a wavelength baseline as wide as
possible, as in the study of \gcs\ in NGC\,5128 by
Jablonka et al. (1996). Although the present wavelength range is 
limited, it is still useful to constrain the reddening values.

We used the \gc\ templates G\,5 to G\,1, ranging from
$\zz=-2.0$\ to nearly-solar metallicity as reference
spectra (Bica 1988). In the case of G\,1, the Bica (1988) template has been
complemented with the M\,31 \gcs\ of comparable
metallicity G\,222 and G\,170 (Jablonka et al. 1992) in order to
further improve the Signal/Noise ratio. For completeness, we also
included the I\,1/I\,2 template, since for some clusters in
Table~\ref{tab_CP}, this possibility cannot be ruled out
(Sect.~5.1). However, we point out that the reddening is not affected
since, in the present spectral domain, the spectral distributions of
I\,1/I\,2 and the old \mp\ \gc\ templates are very
similar; this is illustrated in the top panel of Figure~\ref{fig4}
where we also include an example of reddening determination. In the
bottom panel we show an example of a reddening-corrected \mr\ 
M\,31 \gc\ (Bo\,124) compared to the G\,1 template.

The template adopted in each case is given in column 4 of
Table~\ref{tab_CP}, and the resulting \ebv\ in column 5.

According to Burstein \& Heiles (1984), the foreground (Milky Way)
reddening in the direction of M\,31 is $\ebv_F=0.08$. The results for
the M\,31 cluster sample by Jablonka et al. (1992) suggested a
somewhat lower value $\ebv_F=0.04$. In the present study, the lowest
values and corresponding uncertainties (Table~\ref{tab_CP}) are
compatible with the latter value. For 6 other clusters we derive
values significantly in excess of this limit, which indicates
considerable internal reddening arising probably in dust clouds in the
inner disk of M\,31.

\subsection{A more detailed ranking of metallicities}

\begin{figure} 
\vskip 2.0cm
\hskip -0.0cm
\psfig{figure=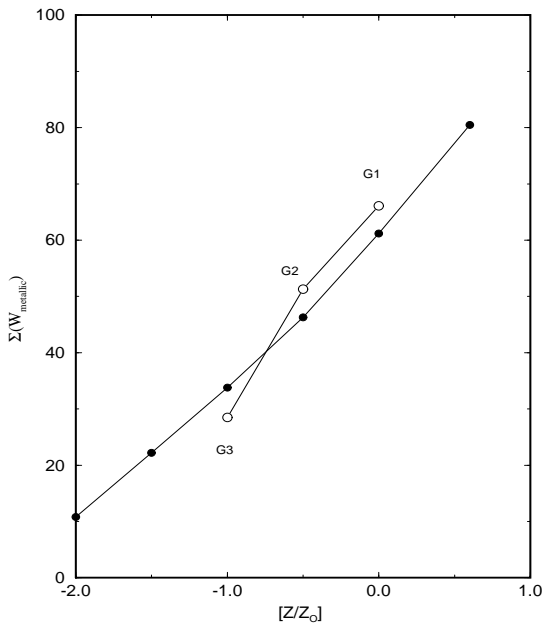,width=8.0cm,height=10.0cm}
\vskip -0.75cm
\caption[]{Sum of EWs of metallic absorption features as a function of
metallicity. Filled circles are according to old clusters in Bica \&
Alloin (1986b); open circles: sum of EWs for the templates G\,1, G\,2
and G\,3 assuming metallicities $\zz=0.0, -0.5$\ and $-1.0$,
respectively, as in Jablonka et al. (1996).}
\label{fig5}
\end{figure}

A more precise ranking of clusters by metallicity may be obtained by
summing the EWs of metallic features, as did Jablonka et al. (1996).
In the present paper we adopt the following absorption windows: \caK\ 
(\#4), Fe\,I (\#10), CN (\#11), Ca\,I (\#12), Fe\,I (\#13), CH G
(\#14) and Fe\,I (\#16). These features are not affected by Balmer
lines and are available for all sample clusters (Table~3). The
resulting sum of the equivalent widths (\sw) is given in column 6 of
Table~4.  The calibration of \sw\ with \zz\ for old clusters in the
grid by Bica \& Alloin (1986b), is displayed in Figure~\ref{fig5}. We
also include in this figure a relation of \sw\ for the templates G\,1,
G\,2 and G\,3, in which they are assumed to have metallicities
$\zz=0.0, -0.5$\ and $-1.0$, respectively, similar to that applied to
NGC\,5128 \gcs\ by Jablonka et al. (1996). Both approaches provide
similar results.

Applying the Bica \& Alloin (1986b) calibration to the M\,31 clusters
(assumed to be genuine \gcs), we obtain the \zz\ values listed in
column 7 of Table~4. Applying the same relation to G\,1, G\,2 and
G\,3, we obtain $\zz=+0.15, -0.33$\ and $-1.23$, respectively.

A fundamental problem in stellar populations is the abundance
calibration of the \mr\ end. For the clusters in common with the
spectroscopic study by Huchra et al. (1991), we show their \fe\ values
in column 8 of Table~4. In general, the same ranking is found, but the
Huchra et al. metallicities are systematically lower.  This question
of calibration has been recently been addressed by means of high
dispersion spectra of individual stars (Barbuy et al. 1992, 1997),
deep colour magnitude diagrams (Ortolani et al. 1995), and integrated
spectra (Santos et al.  1995b), in the case of the key nearly-solar
metallicity \gcs\ NGC\,6553 and NGC\,6528 (which belong to the G\,1
template). The iron abundance \fe\ in such clusters appears to be
somewhat under solar ($\fe\approx-0.2/-0.3$), whereas
[$\alpha$-elements/Fe] are enhanced, resulting in an overall
metallicity $\zz\approx0.0$. In this sense, the present \zz\ 
calibration, which includes $\alpha$-elements, tends to give higher
values but is fundamentally compatible with the \fe\-based scale
employed by Huchra et al. (1991).

We also include in Table~4 (column 9) metallicity estimates from
$(V-K)$\ integrated photometry by Cohen \& Matthews (1994, hereafter
CM94). A comparison of the latter values with those of the present
study (column 7) shows a systematic difference in the sense that CM94
values are lower.  However, the two most \mr\ clusters G\,177 and
G\,174 have spectral absorptions indistinguishable from those of the
most \mr\ galaxy nuclei (Section~6.3). The values estimated by CM94
for these clusters would imply sub-solar metallicities for the nuclei
of giant galaxies. Concerning the metallicity ranking, the agreement
is good between the two studies, with the exception of G\,175, G\,184
and G\,190. G\,175 and G\,184 are considerably reddened (Table~4), and
the higher metallicity values in CM94 can be explained by the fact
that they adopted a constant reddening for their cluster sample.
G\,190 would be a very \mp\ cluster according to their value. However,
the spectral features (Figure~3) are clearly those of a \mr\ cluster;
in particular the G band is strong and the Balmer lines are weak.

\section{Discussion}

\subsection{Metallicity distribution in the bulge}

\begin{figure} 
\vskip 0.0cm
\hskip -0.0cm
\psfig{figure=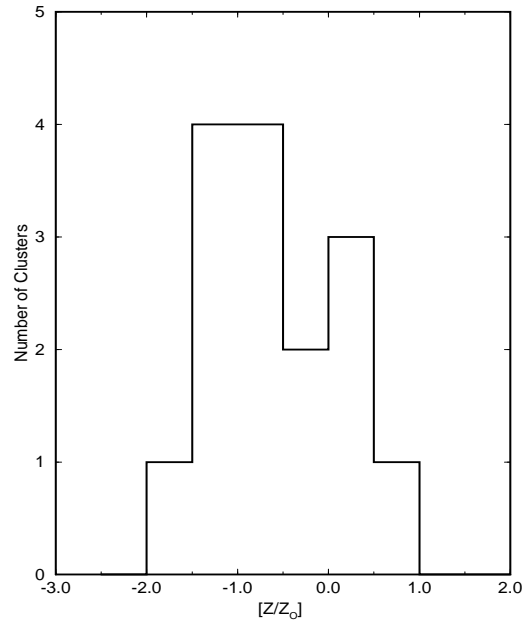,width=8.0cm,height=10.0cm}
\vskip -0.75cm
\caption[]{Metallicity distribution of the M\,31 central clusters,
assuming that the 15 objects in Table~4 are \gcs.}
\label{fig6}
\end{figure}

Assuming that all objects are indeed old globular clusters, we show in
Figure~\ref{fig6} the \zz\ histogram using column 7 of Table~4. The
sample is small, but some interesting features are present. The
distribution is basically Gaussian with an average $\zz=-0.58$\ and
intrinsic $\sigma=0.63$, for this central bulge sample in M\,31.
Recently Barbuy, Bica \& Ortolani (1997) have studied in detail the
sample of 16 Galactic \gcs\ projected within a radius of 5\grau\ of
the Galactic center, based on colour magnitude diagrams; they have
obtained $<\fe>=-0.60$\ with $\sigma=0.61$. Thus, these Galactic and
M\,31 bulge \gc\ samples have similar metallicity distribution. The
Baade's Window sample of K giants studied by McWilliam \& Rich (1994)
has $<\fe>=-0.21$\ with $\sigma=0.49$.  Consequently, both the bulge
\gc\ samples in our Galaxy and in M\,31 seem to be, on the average,
$\approx0.3$\ dex less \mr\ than the bulge stars in our Galaxy
(however we caution that the \gc\ samples are small).

Possible explanations for this difference between \gcs\ and stars are
related to projection effects and/or intrinsic properties. Some
interloping halo clusters are expected in the observed lines of sight
of M\,31 and Galactic central clusters, which might be contaminating
the samples. Alternatively, these differences might reflect different
chemical enrichment histories. However, samples of bulge clusters are
still small and this domain of investigation is still at its early
stages.


Super \mr\ \gcs\ are apparently not a very common phenomenon, since the 
only examples in our sample are G\,174 and G\,177. For more definite results, 
it would be important to increase the spatial coverage of the observations to 
the other quadrants (Fig.~1).

\subsection{Spatial dependence of velocity, reddening and metallicity}

We present in Figure~7 the spatial dependence of metallicity,
reddening and radial velocity for the star cluster sample,
respectively from columns 7 and 5 of Table~4, and an average of
columns 4 and 5 of Table~1. The sample is small and spatially
restricted to one side of the nucleus (Figure~1), but conclusions
can still be drawn.

Clusters of different metallicities are well-mixed. The most reddened
clusters appear superimposed on the galaxy major axis, since there the
disk absorption would be more important. Finally, the velocity
distribution suggests a spatial mixing of clusters which would be in
agreement with the velocity dispersion of a bulge.  The velocity
dispersion of our 16 confirmed clusters is 255 km/sec (unweighted).
To be able to say anything about possible rotation, it would be
necessary to observe the clusters in the other quadrants.

\begin{figure} 
\vskip 2.0cm
\hskip -0.0cm
\psfig{figure=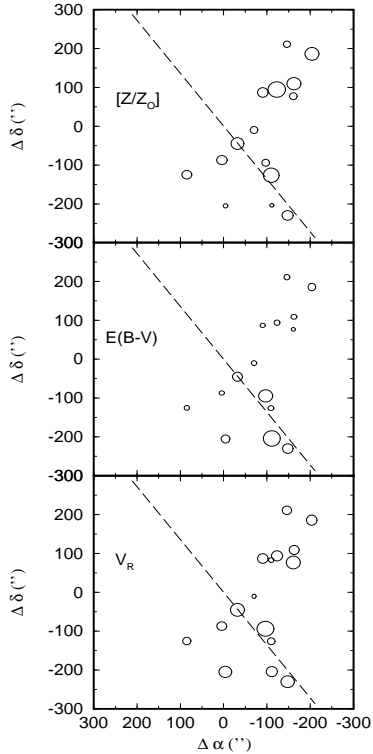,width=8.0cm,height=10.0cm}
\vskip -0.75cm
\caption[]{Spatial dependence of properties of the star cluster
sample. Top panel: metallicity; size of symbols increases with
$[Z/Z_{\sun}]$; middle panel: reddening; size of symbols increases
with \ebv; and bottom panel: radial velocity; size of symbols
increases with V$_R$.}
\label{fig7}
\end{figure}

\subsection{Comparison with the most \mr\ galaxy spectra}

\begin{figure} 
\vskip 0.0cm
\hskip -0.0cm
\psfig{figure=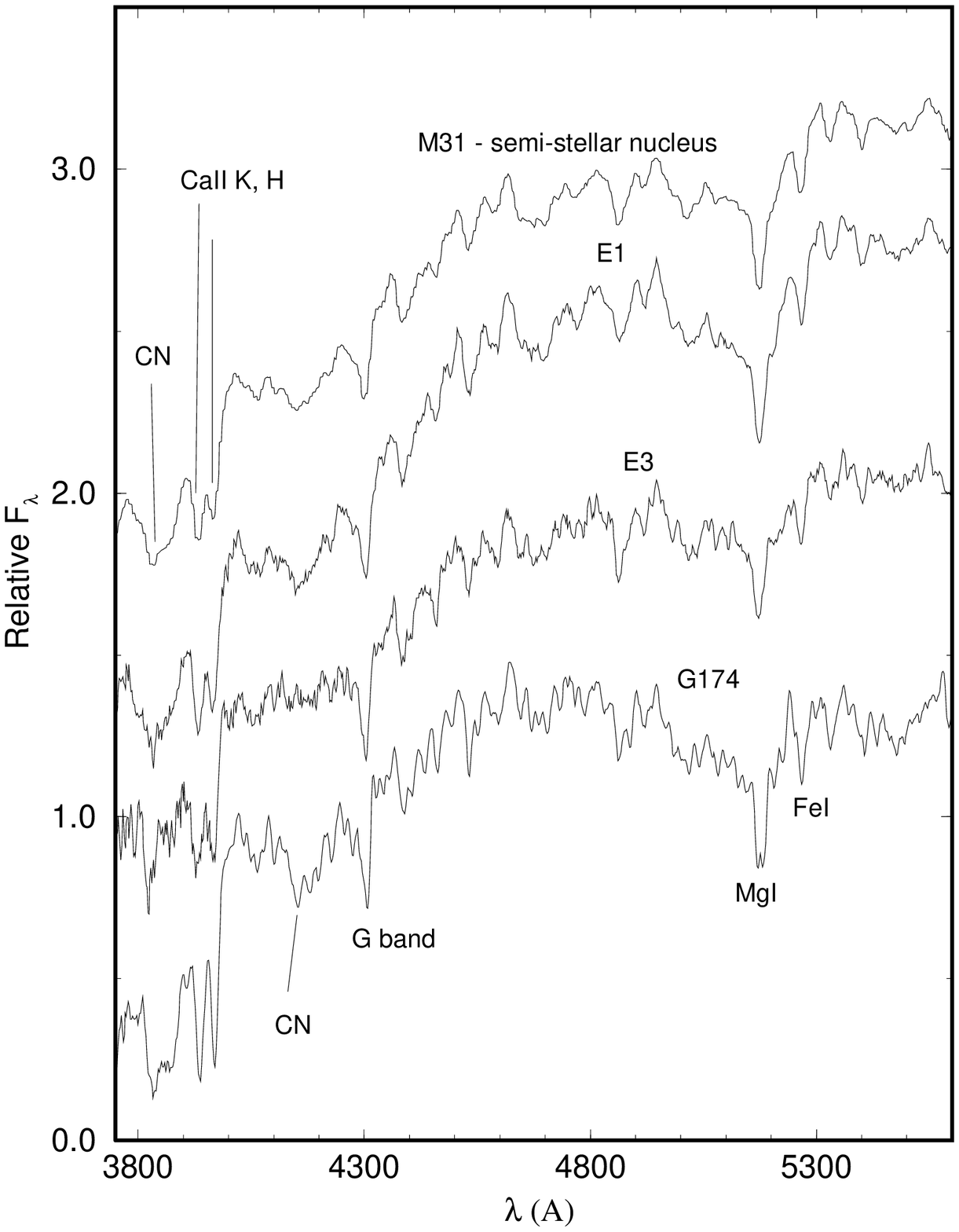,width=8.0cm,height=12.0cm}
\vskip -0.75cm
\caption[]{Comparison of the spectrum of the most \mr\ \gc\ in the 
sample (G\,174) with that of the M\,31 semi-stellar
nucleus (top), the most \mr\ template in nuclei of giant
ellipticals (E\,1), and a less \mr\ template of nuclei of luminous
ellipticals (E\,3). Important metallic features are indicated. Units
as in Figure~2.}
\label{fig8}
\end{figure}

It is important to compare single stellar population (star cluster)
spectral properties with those of composite stellar populations
(galaxies), in view of stellar population synthesis. Bica \& Alloin
(1986a,b and 1987) and Bica (1988) called attention to the fact that
the spectral properties of the nearly-solar metallicity \gcs\ like 
NGC\,6528, NGC\,6553 and NGC\,6440, were almost
comparable to those of giant elliptical galaxies. More recently,
Jablonka et al. (1992) pointed out that the M\,31 bulge \gc\ G\,177 
(also in the present sample) was as strong-lined as the
giant ellipticals. In the present sample, there is a cluster which is
even more strong-lined, G\,174, whose spectrum is compared in
Figure~\ref{fig8} to that of the M\,31 semi-stellar nucleus (Bica, Alloin
\& Schmidt 1990), that of the most \mr\ giant elliptical nuclei (the 
E\,1 template in Bica 1988), and finally, a less \mr\ one, of galaxy
nuclei of somewhat less luminous ellipticals (the E\,3 template in Bica 1988).

The \gc\ G\,174 is clearly more strong-lined than E\,3, and is similar
to those of the most \mr\ known stellar populations, i.e. E\,1 and the
semi-stellar nucleus of M\,31. Notice in particular the comparable
strengths of the blue CN band ($\sim$ 4150\AA) which are among the
most sensitive metallicity indicators, remaining unsaturated in the
high metallicity range. As a consequence, G\,174 appears to have been
formed from the same parent gas as the semi-stellar nucleus, and so it
possibly represents the highest chemical enrichment that galaxies can
attain.

\section{Concluding remarks}

We have carried out spectroscopy of a sample of 19 objects projected on the
bulge of M\,31: 16 were confirmed as star clusters, 2 are Galactic
dwarf stars, and 1 is a high redshift background galaxy. The objects
confirmed as star clusters had been classified as Bologna classes A
and B, whereas the 3 non-clusters were of the low probability class C.

We derived radial velocities, metallicities, ages and reddenings for
our sample clusters. In some cases, especially G\,175, there remains 
an ambiguity between intermediate and old cluster ages, due to 
the limited wavelength range observed. Two clusters are found to be 
super \mr, G\,174 and G\,177. Their features are comparable to 
those in the most \mr\ stellar populations, i.e., nuclei of
giant ellipticals and the semi-stellar nucleus of M\,31 itself. The
metallicity distribution of the sample is compatible with the distribution 
of globulars in the bulge of our Galaxy. Although small, both the M\,31 and
the Galaxy \gc\ samples seem to have an excess of \mp\ 
objects with respect to the stars in the Galactic bulge. Some interloping
halo \gcs\ might be present in both cluster samples. Alternatively, this
might be due to different formation histories.

In the future, it will be important to observe clusters surrounding the 
nucleus in the other quadrants, in order to increase the statistical 
significance of the properties analyzed. In particular this would make it 
possible to check for the existence of rotation in the
cluster system, as well as to 
identify new super \mr\ clusters, imposing constraints on chemical 
evolution models, especially at the high metallicity end.

\begin{acknowledgements}
T.B. thanks the staff at the La Palma Observatory for
hospitality and observational support; E.B. and C.B. acknowledge
financial support from the Brazilian institutions CNPq and FINEP.
\end{acknowledgements}

%
%

\end{document}